\documentclass[10pt,conference]{IEEEtran}
\IEEEoverridecommandlockouts 
\usepackage{graphicx}
\usepackage{amsmath}
\usepackage{amssymb}
\usepackage{setspace}
\usepackage{verbatim}

\usepackage{amsthm}
\usepackage{bbm}
\usepackage{multirow}
\usepackage{psfrag}
\usepackage{booktabs}

\newtheorem{thm}{Theorem}

\newtheorem{lem}[thm]{Lemma}
\newtheorem{defn}[thm]{Definition}
\newtheorem{ex}[thm]{Example}
\newtheorem{assm}[thm]{Assumption}
\newtheorem{alg}[thm]{Algorithm}

\newcommand{\bX}{\mathbf{X}}
\newcommand{\bY}{\mathbf{Y}}
\newcommand{\bZ}{\mathbf{Z}}
\newcommand{\bW}{\mathbf{W}}

\newcommand{\bA}{\mathbf{A}}

\newcommand{\bG}{\mathbf{G}}
\newcommand{\bB}{\mathbf{B}}
\newcommand{\bU}{\mathbf{U}}
\newcommand{\bC}{\mathbf{C}}
\newcommand{\bmu}{\mathbf{\mu}}
\newcommand{\bett}{\mathbf{\eta}}

\newcommand{\tbmu}{\mathbf{\tilde{\mu}}}
\newcommand{\tbett}{\mathbf{\tilde{\eta}}}

\newcommand{\btX}{\mathbf{\tilde{X}}}
\newcommand{\btY}{\mathbf{\tilde{Y}}}

\newcommand{\cN}{\mathcal{N}}

\newcommand{\cC}{\mathcal{C}}

\newcommand{\cS}{\mathcal{S}}

\begin{document}
\title{{\LARGE \vspace{.25 in} A Power Efficient Sensing/Communication Scheme:\\ Joint Source-Channel-Network Coding by Using Compressive Sensing}}

\author{
Soheil Feizi\\MIT\\Email: sfeizi@mit.edu \and Muriel M\'edard\\ MIT\\Email: medard@mit.edu 

\thanks{This material is based upon work supported by AFOSR under award No. 016974-002.}
}

\maketitle
%\doublespacing

\begin{abstract}

We propose a joint source-channel-network coding scheme, based on compressive sensing principles, for wireless networks with AWGN channels (that may include multiple access and broadcast), with sources exhibiting temporal and spatial dependencies. Our goal is to provide a reconstruction of sources within an allowed distortion level at each receiver. We perform joint source-channel coding at each source by randomly projecting source values to a lower dimensional space. We consider sources that satisfy the restricted eigenvalue (RE) condition as well as more general sources for which the randomness of the network allows a mapping to lower dimensional spaces. Our approach relies on using analog random linear network coding. The receiver uses compressive sensing decoders to reconstruct sources. Our key insight is the fact that, compressive sensing and analog network coding both preserve the source characteristics required for compressive sensing decoding.
 
\end{abstract}

\section{Introduction}

The power budget and communication rates are two main limitations in wireless network applications. Information theory focuses much on minimizing transmission rates, but the complexity of coding processes. For example, in point-to-point channels, equivalence theory allows us to separate the problem of channel coding from that of joint network/source coding \cite{kornerbook}. However, any approach that separates source coding from channel coding may suffer from potential inefficiencies since we reduce redundancy for source coding but later reintroduce it for channel coding. Moreover, even if we allow separate channel coding from source/network coding (which is necessary because of lack of separation \cite{ram}), the known schemes have generally high complexity inherited from the general Slepian-Wolf problem \cite{sw}. 

  \begin{figure}[t]
	\centering
    \includegraphics[width=8.5cm]{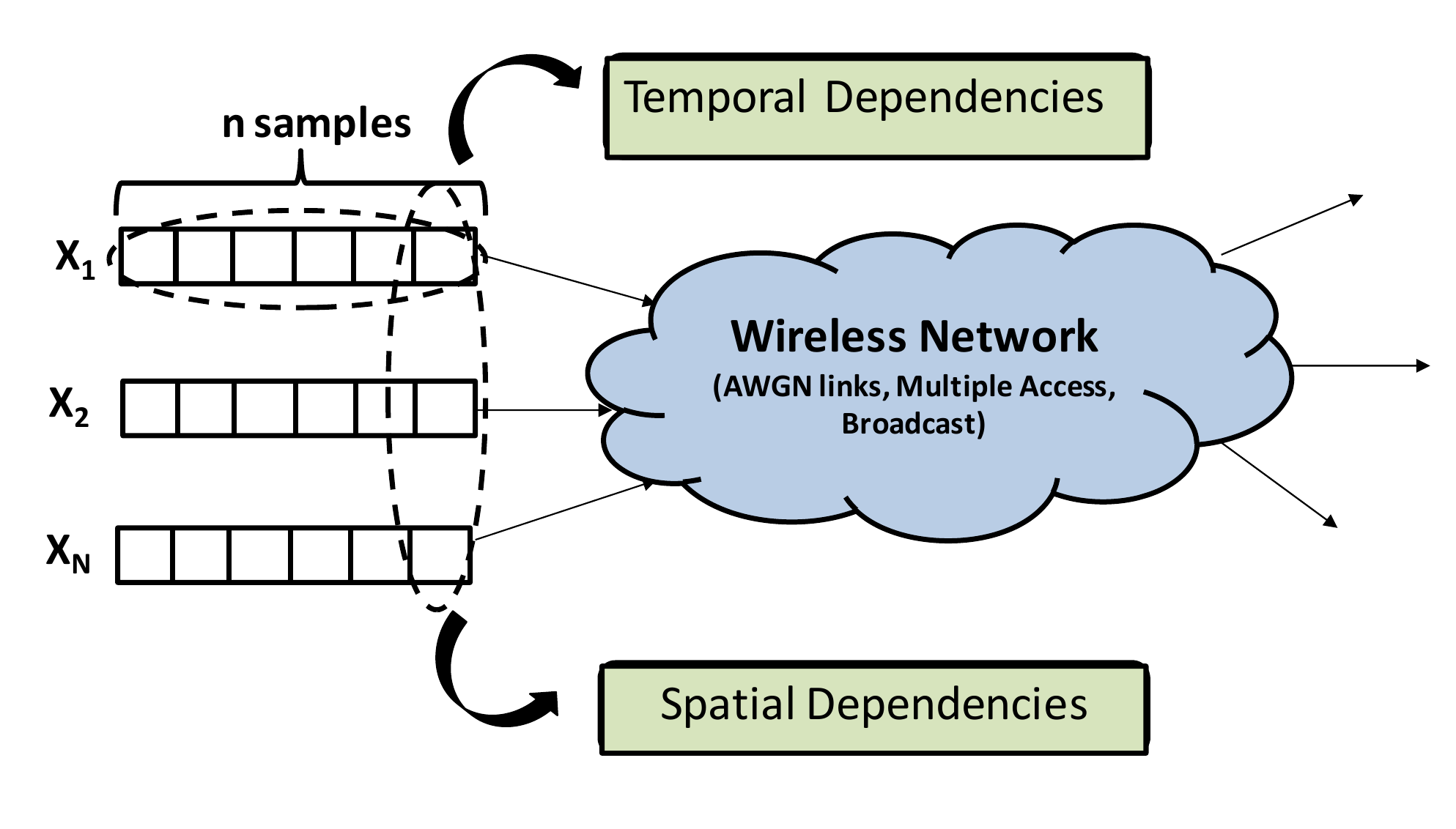}
    \caption{The considered problem setup where sources have temporal and spatial dependencies, and receivers desire to reconstruct sources within an allowed distortion level.}
    \label{fig:framework}
  \end{figure}

Compressive sensing (\cite{candes}, \cite{candes-noiseless}, \cite{donoho}) can be used to perform channel coding (\cite{feizi-CS}) as well as source coding, even for distributed sources (\cite{feizi-CS}). In both of these cases, the performance of the compressive sensing approach can be shown to be close to that of the information theoretic optimum. The fact that, a single approach may successfully be applied for channel coding and source coding renders it an attractive candidate for use in a joint problem. Moreover, the fact that, linear analog network coding is asymptotically optimal in the high SNR regime \cite{anc}, together with the fact that, compressive sensing-based channel and source techniques are also linear, provides a compelling reason for considering a compressive sensing-based joint source-channel-network coding. 

Here, we propose a joint source-channel-network coding scheme, based on compressive sensing principles, for wireless networks with AWGN channels (that may include multiple access and broadcast). We assume that, sources exhibit temporal and spatial dependencies. We consider a multicast wireless network with multiple sources and receivers (see Figure \ref{fig:framework}). Our goal is to provide a reconstruction of all sources within an allowed distortion level at all receivers in a low-complexity manner. A key idea of our proposed scheme is to reshape inherent redundancies among source samples (temporal or spatial) in a joint source-channel-network coding scheme. Our scheme does not rely on the distribution knowledge of sources. Also, it provides an attractive relationship between the number of uses of the network and the reconstruction error.

\begin{figure}[t]
	\centering
    \includegraphics[width=8.5cm]{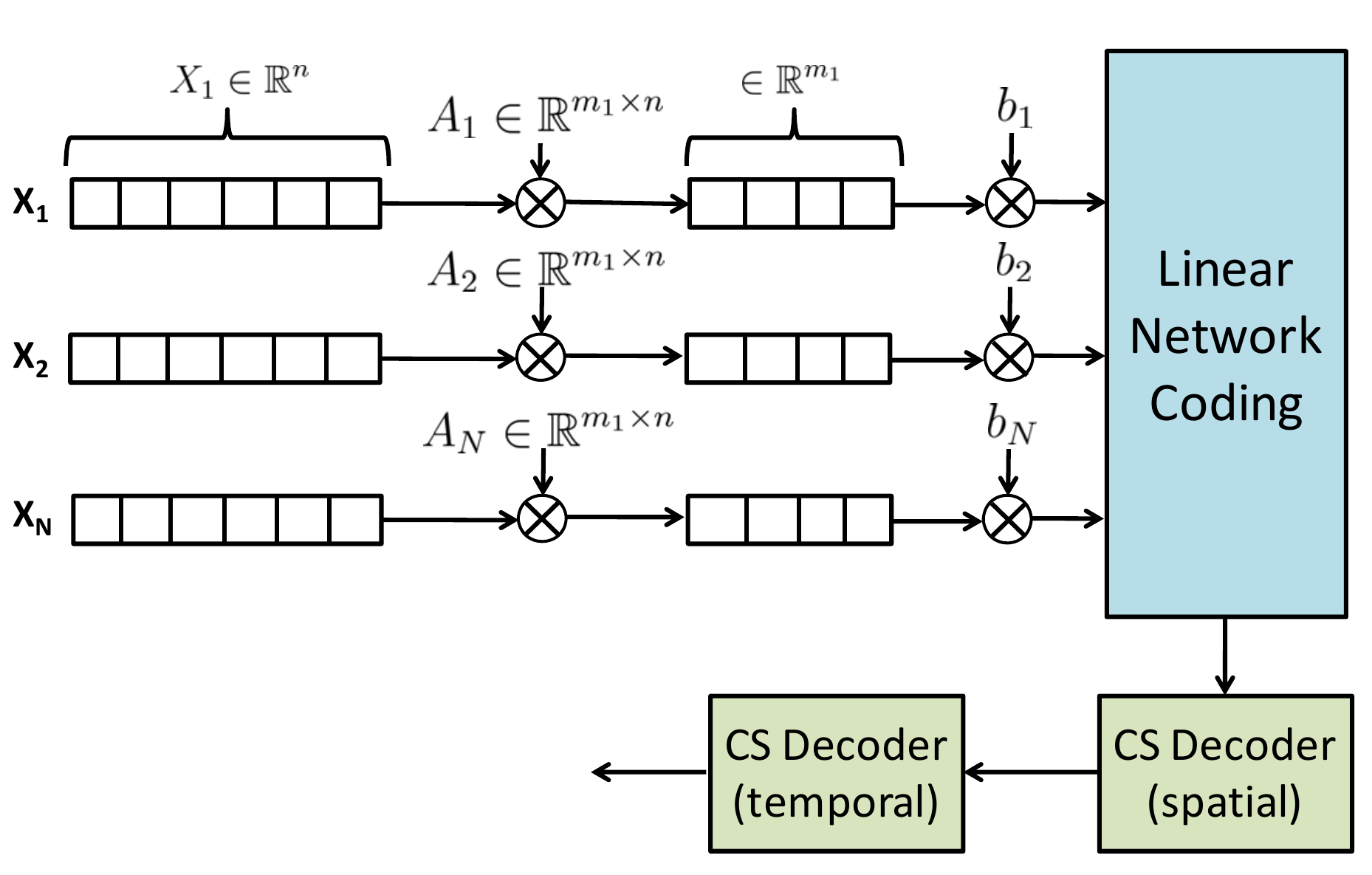}
    \caption{A schematic view of our proposed joint source-channel-network coding scheme. Each source projects its high dimensional signal to a lower dimensional space by using a random projection matrix followed by a pre-coding on-off module. Intermediate nodes perform linear network coding. Each receiver uses compressive sensing decoders (LASSO) to reconstruct source signals.}
    \label{fig:alg}
  \end{figure}

A schematic view of our proposed framework is illustrated in Figure \ref{fig:alg}. In this framework, we perform joint source-channel coding at each source by randomly projecting source values to a lower dimensional space. We consider sources that spatially satisfy the restricted eigenvalue (RE) condition and may use sparse networks as well as more general sources, which use the randomness of the network to allow mapping to lower dimensional spaces. Our scheme relies on using random linear network coding in the real field. The receiver uses compressive sensing decoders (spatial and temporal ones) to reconstruct source signals.

Our main insight is that, the main characteristics required for operating compressive sensing are preserved under very general linear operations, which may themselves be arise from cascade of linear transformations in a network. We can therefore cascade compressive sensing techniques as well as linear operations tied to coding in the interior of the network. This preservation of the main characteristics of the sources under linear transformations lies at the core of both compressive sensing and analog network coding. This work constitutes a step in bringing together these two techniques, which have natural similarities.  

This paper is organized as follows. In Section \ref{sec:background}, we review some prior results in compressive sensing and random projections. We propose our scheme in Section \ref{sec:main} and characterize its performance. Proofs are presented in Section \ref{sec:proofs}. We conclude the paper in Section \ref{sec:conc}.

\section{Compressive Sensing Background}\label{sec:background}
 
In this section, we review some prior results in compressive sensing. Let $\bY\in\mathbb{R}^N$ be an unknown signal vector. We say this signal is $k$-sparse if $k$ of its coordinates are non-zero and the rest are zero. Note that, the assumption of having $N-k$ components of $\bY$ to be exactly zero may seem unrealistic for practical cases. This assumption is called a hard sparsity assumption. Weakly sparse models are also considered in the compressive sensing literature \cite{weak-sparse}. Roughly, a vector $\bY$ is weakly sparse if it can be approximated by a sparse vector. In this paper, we consider hard sparsity models although all discussions can be extended to weakly sparse signals.

 Let $G\in\mathbb{R}^{m\times N}$ be a measurement matrix where $m<<N$. The observation vector $\bZ$ is defined by the following linear model:  

\begin{equation}
\bZ=\bG\bY+\bW,\nonumber
\end{equation}

where $\bW\in\mathbb{R}^m$ is a noise vector. We assume $w_i\sim \cN(0,\sigma^2)$, where $\cN(0,\sigma^2)$ represents a Gaussian random variable with mean zero and variance $\sigma^2$.   

By obtaining observations $\bZ$ and knowing measurement matrix $\bG$, the goal is to reconstruct $\btY$ such that
\begin{equation}
\|\bY-\btY\|_{l_2}^2\leq D,\nonumber
\end{equation}

where $D$ is a distortion allowance threshold. If the signal $\bY$ is a sparse signal (i.e., $k<<N$) and the measurement matrix satisfies the \textit{Restricted Eigenvalue (RE) Condition} \cite{RE} (which will be explained later) or more restrictively, \textit{Restricted Isometry Property} (\cite{candes-rip}), then having $m$ measurements where $m<<N$ is sufficient to recover the original sparse vector. In fact, $\btY$ is a solution for the following convex optimization called LASSO (\cite{lasso}): 

\begin{align}\label{eq:lasso}
\min_{\bY}\quad& \frac{1}{2m}\|\bZ-\bG\bY\|_{l_2}^2+\xi \|\bY\|_{l_1},
\end{align}

where $\|.\|_{l_p}$ represents the $l_p$ norm of a vector and $\xi>0$ is a regularization parameter. If $\btY$ is a solution of this optimization, then, \cite{RE} shows that,

\begin{equation}\label{eq:error-bound}
\|\bY-\btY\|_{l_2}^2\leq \frac{\delta}{\gamma^2} \sigma^2 \frac{k\log(N)}{m}
\end{equation}

with high probability, where $\delta$ is a constant and $\gamma$ is a parameter related to the restricted eigenvalue condition of the matrix $\bG$. 

  \begin{figure}[t]
	\centering
    \includegraphics[width=8.5cm]{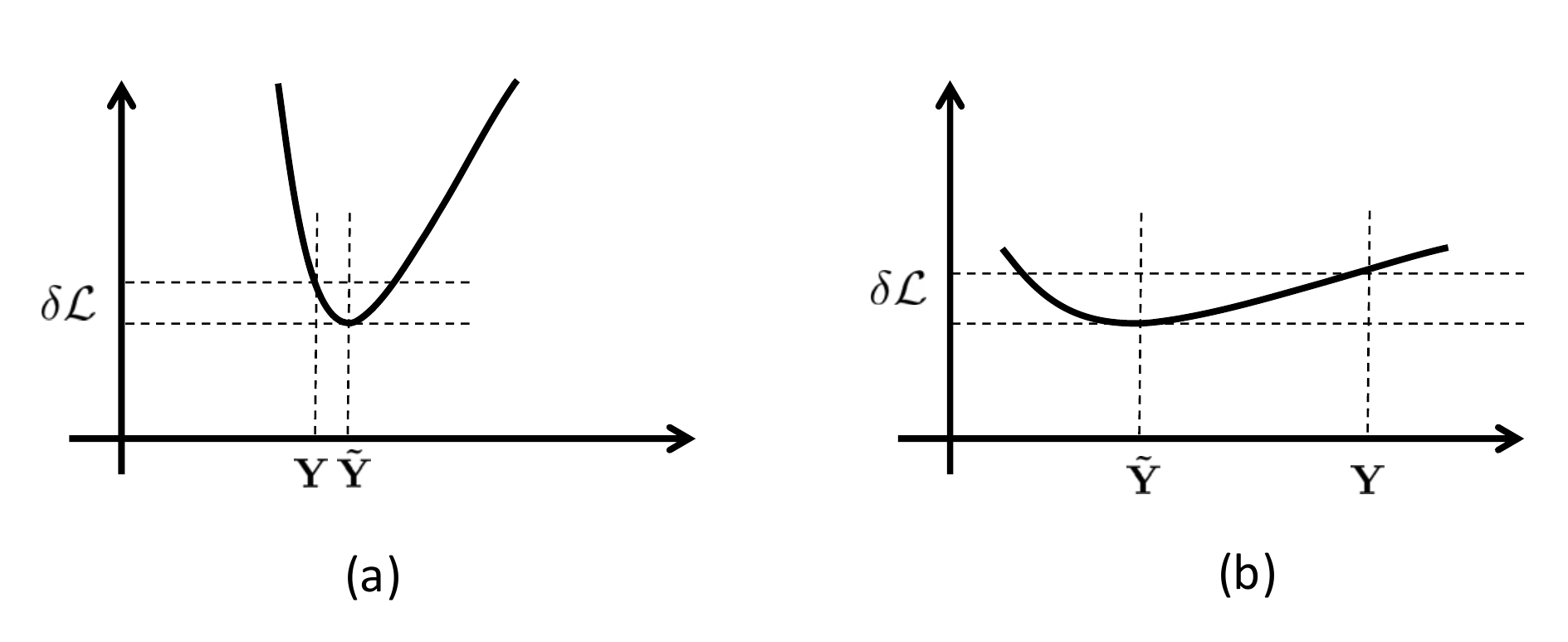}
    \caption{To control the reconstruction error, the loss function should have high curvature around the optimal solution of LASSO (equation \ref{eq:lasso}) along the directions of $\cC(\cS;\alpha)$.}
    \label{fig:RE}
  \end{figure}

Now, we define the restricted eigenvalue condition (RE) for the measurement matrix $\bG\in\mathbb{R}^{m\times N}$ (\cite{RE}). Suppose $\cS$ is the support of a sparse vector $\bY$ (i.e., $\cS$ is the set of non-zero indices of $\bY$ where $|\cS|=k$). $\cS^c$ is the complement set of $\cS$. We define the subset
\begin{equation}
\cC(\cS;\alpha)=\{\bY\in\mathbb{R}^{N}: \|\bY_{\cS^c}\|_{l_1} \leq \alpha \|\bY_{\cS}\|_{l_1} \}.\nonumber
\end{equation}

This corresponds to the cone of vectors where $l_1$ norm on their support dominates the $l_1$ norm off the support. For $\alpha=1$, this cone is the same as the one used in restricted null-space property (\cite{RNS}).

\begin{defn}
The matrix $\bG$ satisfies the restricted eigenvalue (RE) condition over $\cS$ with parameter $(\gamma,\alpha)$ if
\begin{equation}
\frac{\frac{1}{m}\|\bG\bY\|_{l_2}^2}{\|\bY\|_{l_2}^2}\geq \gamma > 0
\end{equation}
for all $\bY\in \cC(\cS;\alpha)-\{0\}$.
\end{defn}

The proof of equation ($\ref{eq:error-bound}$) can be found in reference (\cite{RE}). By abuse of notation, we say a square matrix satisfies the restricted eigenvalue condition with parameter $(\gamma,\alpha)$ if all of its sub-matrices with $|\cS|$ rows satisfies the RE condition with parameter $(\gamma,\alpha)$. Also, roughly we say a matrix satisfy the RE condition, if $\gamma$ is large enough. In the following, we develop some intuitions about the compressive sensing decoder of equation (\ref{eq:lasso}) and the RE condition.

The restricted eigenvalue condition bounds away eigenvalues of the matrix $\frac{1}{m}\bG^{T}\bG$ from zero in the direction of the considered cone. Note that, to be able to control error of LASSO in equation (\ref{eq:lasso}), around the optimal point, the loss function should have high curvature to translate small $l_2$ loss to small error in $\bY$ (see Figure \ref{fig:RE}). For this loss function, Hessian (curvature) can be calculated as $\frac{1}{m}\bG^{T}\bG$. Therefore, we want to bound eigenvalues of $\frac{1}{m}\bG^{T}\bG$ away from zero to have high curvature of the loss function. However, since $\bG$ is a fat matrix (i.e., $m<<N$), some of its eigenvalues are zero. That is why we require to have eigenvalues away from zero for a subset $\cC(\cS;\alpha)$ of vectors. This subset is well-aligned with the directions of the Hessian in which loss is observed (\cite{RE}).  

It is shown in different references that random matrices satisfy the RE condition. Reference \cite{bar-rip} shows a connection between compressive sensing, $n$-widths, and the Johnson-Lindenstrauss Lemma. Through this connection, \cite{database} proposes some data-base friendly matrices (binary matrices) satisfying the RE condition. 

Our goal is in this paper is to use random projection principles to obtain a power-rate efficient sensing-communication scheme. New results and bounds in compressive sensing can potentially improve results of this paper. One can find more details on compressive sensing results in \cite{candes}, \cite{candes-noiseless}, \cite{donoho}, etc. 

In the next section, we present main results of this paper.

\section{Main Results}\label{sec:main}

In this section, we present main results of this paper. First, we present a Lemma demonstrating how the RE condition changes when matrices are cascaded (multiplied to each other). Then, after introducing the problem setup and some notations, we present our proposed joint source-channel-network coding scheme. Finally, we characterize the performance of this proposed scheme. Proofs are presented in the next section.

\subsection{Background Results}
  
In this section, we present a Lemma which is central to implementing the cascading linear transformations which form the basis of our approach.

\begin{lem}[Cascading Lemma]\label{lem:cascade}
Suppose $\bG\in\mathbb{R}^{m\times N}$ satisfies the RE condition with parameter $(\gamma,\alpha)$. Say $\bC_1\in\mathbb{R}^{m\times m}$ and $\bC_2\in\mathbb{R}^{N\times N}$. Suppose minimum absolute eigenvalues of $\bC_1$ and $\bC_2$ are $\lambda_1$ and $\lambda_2$, respectively. Also, say for $\bY\in\cC(\cS;\alpha)$, $\bY\bC_2\in\cC(\cS;\alpha)$. Then, $\bC_1\bG$ and $\bG\bC_2$ satisfy the RE condition with parameters $(\gamma \lambda_1^2,\alpha)$ and  $(\gamma \lambda_2^2,\alpha)$, respectively.
\end{lem}

The proof of this Lemma is presented in Section \ref{sec:proofs}. Intuitively, this lemma says, if $\bG$, which satisfies the RE condition, is cascaded with other matrices whose eigenvalues are bounded away from zero, the overall matrix would satisfy the RE condition.

\subsection{The problem Setup and Notations}

We consider a wireless network with $N$ sources and $l$ receivers. Each source has $n$ correlated samples to transmit to all receivers (see Figure \ref{fig:framework}). Also, we assume that, there are temporal and spatial correlations among source samples. Vector $\bX_i\in\mathbb{R}^n$ represents samples of source $i$ and $\bX\in\mathbb{R}^{N\times n}$ is a matrix of all source samples. $x_i^t$ represents a sample of source $i$ at time $t$.

We have the following assumptions on source samples capturing their temporal and spatial redundancies:

\begin{assm}\label{assm:cor}
Sources have temporal and spatial dependencies:

1) Each source has temporal dependency: there exists a matrix $\Phi\in\mathbb{R}^{n\times n}$ such that $\bX_i=\Phi \bett_i$, where $\bett_i$ is $k_1$ sparse and $k_1<<n$.

2) Source samples have spatial dependency: suppose $\bY\in\mathbb{R}^{N}$ is a vector containing $y_j=\bA_j^T \bX_j$ for some vector $\bA_j\in\mathbb{R}^{n}$, for $1\leq j\leq N$. There exists a matrix $\Psi\in\mathbb{R}^{N\times N}$ such that, $\bY=\Psi \bmu$, where $\bmu$ is $k_2$ sparse, and $k_2<<N$.
\end{assm}

Note that, matrices $\Phi$ and $\Psi$ are not required to be known at sources, but at receivers only. This is an important issue in hardware implementations of sensors (\cite{sensor}). We will show that, our proposed scheme is generic.

Each receiver desires to obtain all sources within an allowed distortion level $D$. In other words, for all $1\leq i\leq N$ and all receivers, we should have
\begin{equation}\label{eq:dist-req}
\frac{1}{n}\|\bX_i-\btX_i\|_{l_2}^2\leq D,
\end{equation}

with high probability, where $\btX_i$ is the reconstructed signal of source $i$.

Links of the network are AWGN channels. Nodes in the network are allowed to perform linear network coding in the real field (see analog network coding references \cite{anc-dina}, \cite{anc}, etc.). We assume that, the network can deliver $m$ linear combinations of source samples to each receiver at each use of the network. In other words, at each use of the network, a receiver obtains $\bZ\in\mathbb{R}^{m}$ where, $\bZ=\bG\bY+\bW$. $\bG\in\mathbb{R}^{m\times N}$ is the network coding matrix, $\bY\in\mathbb{R}^{N}$ is a vector that sources transmit at that time (i.e., source $i$ transmits $y_i$). $\bW$ is the noise vector where we assume, $w_i\sim N(0,\sigma^2)$, for all $1\leq i\leq m$. The quantity $m$ can be viewed as a real field version of the min-cut rate of the network defined in the finite field (\cite{ahl2000}, \cite{medard2003}). We assume that, the number of source samples (i.e., $n\times N$) is much larger than $m$. We define $\cC_{\mbox{use}}$ as the number of network uses to deliver all source samples to receivers so that, the reconstruction distortion requirement (equation \ref{eq:dist-req}) holds for each receiver. We characterize $\cC_{\mbox{use}}$ for our proposed sensing-communication framework and compare it with a naive information theoretic approach.   

\subsection{Joint Source-Channel-Network Coding Scheme}    

In this section, we present our proposed joint source-channel-network coding scheme based on compressive sensing principles. We also characterize the rate-distortion performance of our proposed scheme in Theorem \ref{thm:main}. 

Our proposed sensing-communication scheme can be performed in the following steps, as illustrated in Figure \ref{fig:alg}. We will explain each step in more detail.

\begin{alg}\label{alg:CSJSC}
Our proposed joint source-channel-network coding can be performed in following steps. 

\begin{itemize}
\item {\bf Step 1 (Temporal Pre-Coding):} Each source projects its $n$ dimensional signal to a lower dimensional space by using random projection matrices.
\item {\bf Step 2 (Spatial Pre-Coding):} Each source $i$ transmits with certain probability $b_i$, at each time.
\item {\bf Step 3 (Network Coding):} Nodes in the network perform analog random linear network coding.
\item {\bf Step 4 (Decoding):} Each receiver uses compressive sensing decoders (LASSO) to reconstruct source signals.
\end{itemize}
\end{alg}

These steps are shown in a schematic way in Figure \ref{fig:alg}. In the following, we explain each step in more details:

{\bf Step 1 (Temporal Pre-Coding):} Each source $i$ projects its sample vector $\bX_i$ to a lower dimensional space, by using a matrix multiplication. That is, $\bY_i=\bA_i\bX_i$, where $\bX_i\in\mathbb{R}^{n}$ and $\bA_i\in\mathbb{R}^{m_1\times n}$. Note that, $m_1<<n$, and $m_1$ is chosen according to Theorem \ref{thm:main} to control the reconstruction error. Matrix $\bA_i$ is a random matrix that satisfies the RE condition with parameter $(\gamma_1,\alpha)$.   

{\bf Step 2 (Spatial Pre-Coding):} At time $t$, each source transmits $y_i^t$ (the $t$-th component of $\bY_i$) with probability $b_i$. In the next step, we will show how by choosing appropriate $b_i$, the transmission power can be decreased further.

{\bf Step 3 (Network Coding):} Nodes in the network perform random linear network coding over the real field. At each use of the network, it can deliver $m$ linear combinations of transmitted samples to receivers. Therefore, $m_2/m$ network transmissions are required to gather $m_2$ linear combinations of $y_i^t$ at each receiver, with high probability. Note that, $m_2<<N$ is chosen according to Theorem \ref{thm:main} to control the reconstruction error. By abuse of notation, we drop superscript $t$. Hence, a receiver obtains:
\begin{equation}
\bZ=\bG\bB\bY+\bW=\bG\bB\Psi\bmu+\bW,
\end{equation}

where $\bY\in\mathbb{R}^{N}$ contains $y_i$ for $1\leq i\leq N$,  $\bB\in\mathbb{R}^{N\times N}$ is a diagonal matrix whose $i$-th diagonal element is one with probability $b_i$, and zero otherwise. $\bG\in\mathbb{R}^{m_2\times N}$ is the network coding matrix for that receiver. Note that, by using Assumption \ref{assm:cor}, $\bY=\Psi\bmu$, where $\bmu$ is $k_2$ sparse.

To be able to use compressive sensing principles and decoders, the matrix $\bG\bB\Psi$ should satisfy the RE condition explained in \ref{sec:background}. We consider two cases: (1) $\Psi$ satisfies the RE condition, (2) $\Psi$ does not satisfy the RE condition.

In case (1), since $\Psi$ satisfies the RE condition, by using Lemma \ref{lem:cascade}, $\bG\bB\Psi$ would satisfy the RE condition. Since the rank of $\bG\bB\Psi$ needs to be $m_2$ with high probability, we choose $b_i=1$ with probability $m_2/N$, and $b_i=0$ with probability $1-m_2/N$. In other words, only a fraction of sources are transmitting. Also, by performing random linear network coding over the real field, rank of the matrix $\bG$ would be $m_2$, with high probability (\cite{random}). Hence, for a general network structure, receivers can use compressive sensing decoders to reconstruct source signals within an allowed distortion level.   

In case (2) where $\Psi$ does not satisfy the RE condition, we choose $b_i=1$ for all $i$ (i.e., all sources are transmitting). Hence, $\bB$ is the identity matrix. In this case, if $\bG$ satisfies the RE condition, by using Lemma \ref{lem:cascade}, $\bG\Psi$ would satisfy the RE condition, and therefore, receivers can use compressive sensing decoders to reconstruct source signals.

Intuitively, to have $\bG$ satisfy the RE condition, the network should be dense to randomly mix source samples. For networks that are not sufficiently dense, such as tree networks, $\bG$ never satisfies the RE condition. The following example presents a family of networks for which $\bG$ can be designed to satisfy the RE condition. 
 
  \begin{figure}[t]
	\centering
    \includegraphics[width=7cm]{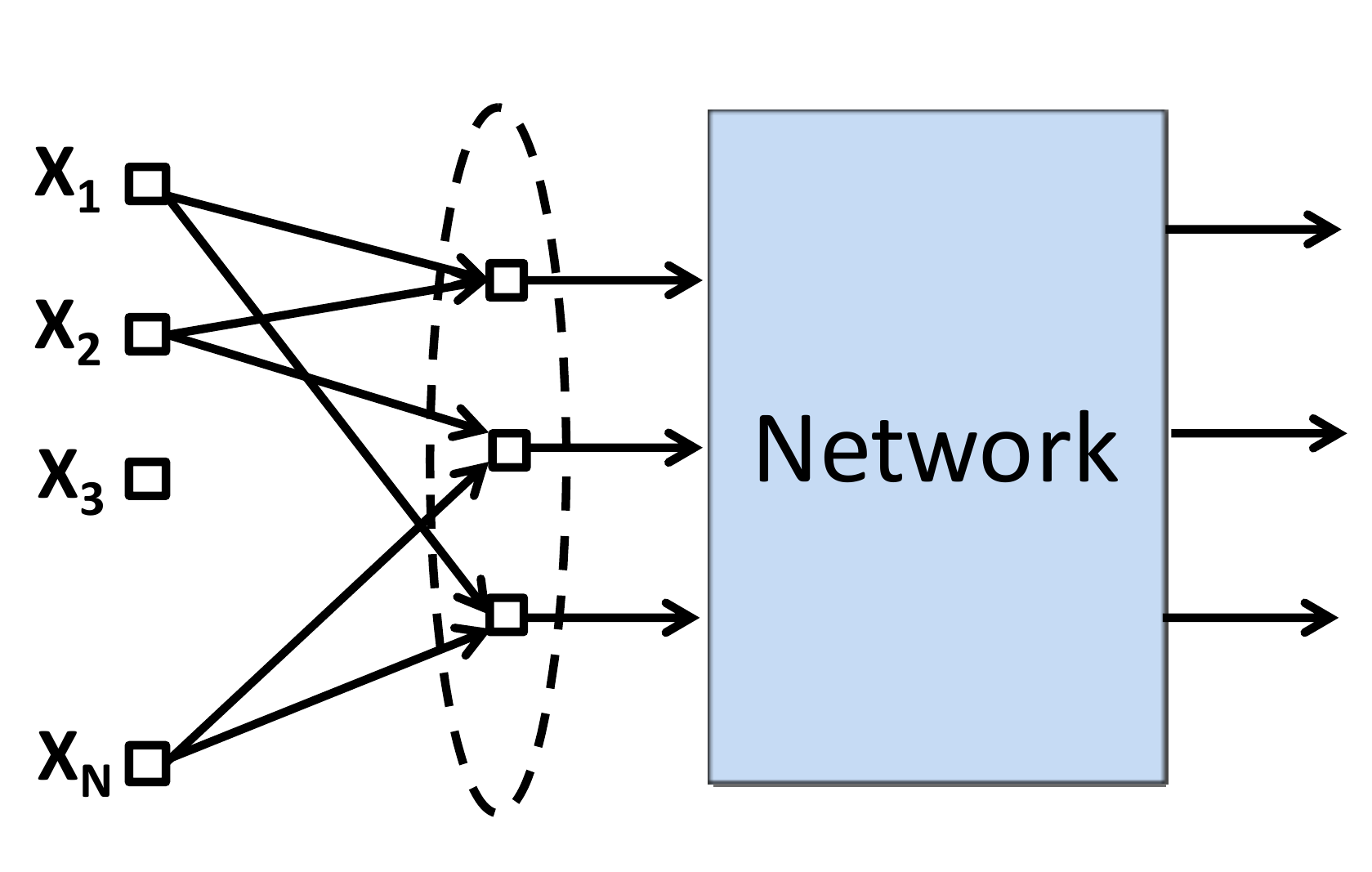}
    \caption{An example of a network topology where the network coding matrix can be designed to satisfy the RE condition. Hence, the network can transform sources with spatial dependencies which do not satisfy the RE condition, to a proper lower dimensional space.}
    \label{fig:re-net}
  \end{figure}

\begin{ex}\label{ex:re-network}
Consider the network topology with $N$ sources depicted in Figure \ref{fig:re-net}. Sources are connected randomly to $m$ intermediate nodes. Expected in-degree of each of these intermediate nodes is greater than or equal to $N/3$ (i.e., high in-degree nodes). These intermediate nodes are connected to another network. The transform matrix of this network (i.e., $\bG\in\mathbb{R}^{m\times N}$) can be decoupled into two matrices: the one for the first stage of the network $\bG_1\in\mathbb{R}^{m\times N}$ and, the one for the rest, $\bG_2\in\mathbb{R}^{m\times m}$. That is, $\bG=\bG_2\bG_1$. If sources choose network coding coefficients uniformly from $\{-1,1\}$, as shown in reference \cite{database}, the matrix $\bG_1$ satisfies the RE condition. By performing random linear network coding, the rank of $\bG_2$ is $m$ with high probability. Therefore, by using Lemma \ref{lem:cascade}, $\bG$ satisfies the RE condition.   
\end{ex}

{\bf Step 4 (Decoding):} Each receiver first uses a compressive sensing decoder as in equation (\ref{eq:lasso}) to reconstruct $\bY^{t}$ for each $1\leq t\leq m_1$ as follows:

\begin{align}\label{eq:lasso-decoder-1}
\min_{\bmu}\quad& \frac{1}{2m_2}\|\bZ-\bG\bB\Psi\bmu\|_{l_2}^2+\xi \|\bmu\|_{l_1}.
\end{align}

Therefore, for each source $i$, $\btY_i$ is obtained where $\btY_i=\bA_i \Phi \bX_i^{s}+\bU_i$. $\bU_i$ is the error occurred in the reconstruction of $\bY^{t}$. Then, another compressive sensing decoder is used to reconstruct $\btX_i$:

\begin{align}\label{eq:lasso-decoder-2}
\min_{\bett_i}\quad& \frac{1}{2m_1}\|\btY_i-\bA_i\Phi\bett_i\|_{l_2}^2+\xi \|\bett_i\|_{l_1}.
\end{align}

The number of network uses of this proposed sensing-communication scheme is equal to $\cC_{\mbox{use}}=\frac{m_1m_2}{m}$. 

$\bA_i\Phi$ satisfies the RE condition with parameters $(\gamma_1\lambda_1^2,\alpha)$ when $\bA_i$ satisfies the RE condition with parameters $(\gamma_1,\alpha)$ and $\lambda_1$ is the minimum absolute eigenvalue of the matrix $\Phi$ (according to Lemma \ref{lem:cascade}). The matrix $\bG\bB\Psi$ satisfies the RE condition with parameters $(\gamma_2\lambda_2^2,\alpha)$ (i.e., in case 1, matrix $\bB\Psi_{\cS}$ satisfies the RE condition with parameter $(\gamma_2,\alpha)$ and $\lambda_2$ is the minimum absolute eigenvalue of $\bG^{\cS}$ where $\bG^{\cS}$ contains columns of $\bG$ corresponding to active sources. In case 2, $\bG$ satisfies the RE condition with parameters $(\gamma_2,\alpha)$ and $\lambda_1$ is the minimum absolute eigenvalue of the matrix $\Psi$). Let $\lambda_3$ and $\lambda_4$ be maximum eigenvalues of matrices $\Psi$ and $\Phi$, respectively. By using these parameters, define the following constant:

\begin{equation}\label{eq:const}
c=\delta\frac{(\lambda_3\lambda_4)^2}{(\gamma_1\gamma_2)^2(\lambda_1\lambda_2)^4},
\end{equation}

where $\delta$ is a constant. The following Theorem illustrates how to choose $m_1$ and $m_2$ in the proposed joint source-channel-network coding scheme to reconstruct each source signal within an allowed distortion level:  

\begin{thm}\label{thm:main}
A joint source-channel-network coding scheme of Algorithm \ref{alg:CSJSC} satisfies the distortion requirement of equation (\ref{eq:dist-req}) with high probability if
\begin{equation}
\cC_{\mbox{use}}= c \frac{k_1k_2\log(n)\log(N)}{m} \frac{\sigma^2}{D},
\end{equation}

where $k_1<<n$ and $k_2<<N$ represent temporal and spatial sparsity of sources, $\sigma^2$ is the noise variance, $D$ is the allowed reconstruction distortion level, and $c$ is a constant defined in equation (\ref{eq:const}). 
\end{thm}

The proof of this Theorem is presented in Section \ref{sec:proofs}. This theorem provides a way of choosing parameters $m_1$ and $m_2$ in Algorithm \ref{alg:CSJSC}, where each source signal can be reconstructed within an allowed distortion level (equation (\ref{eq:dist-req})). 

\textit{Remarks:}
\begin{itemize}
\item Our proposed scheme is efficient: Sources use naturally occurring redundancies to perform joint source-channel coding. Therefore, there is no need of performing source coding and channel coding separately which would cause inefficiencies. Also, in the case that $\Psi$ satisfies the RE condition, only a fraction of sources are required to transmit, which provides additional power saving in transmissions. 

\item Our proposed scheme is broadly structure independent: Matrices $\Phi$ and $\Psi$ are not required to be known at sources, but at receivers. It is important in the hardware implementation of sensors for different applications where measured signals can have different correlation structures (\cite{sensor}). 

\item Our proposed scheme has a continuous rate-distortion performance: as it can be seen from Theorem \ref{thm:main}, if the noise variance $\sigma^2$ increases, by using the same sensing-communication scheme, the reconstruction distortion would increase in a continuous manner (i.e., there is no performance drop off).  In fact, our proposed scheme can be viewed as a \textit{noise modulating} sensing-communication scheme. 

\item Our proposed scheme has low complexity decoders. Unlike Slepian-Wolf (\cite{sw}) decoders of information theoretic coding schemes, decoders used in our scheme are convex optimizations and therefore have lower complexity. 

\item It is insightful to compare the rate-distortion performance of our scheme with a naive information theoretic scheme ignoring source correlations. Following the rate-distortion Theorem (\cite{kornerbook}), to represent a Gaussian random variable $\cN(0,\sigma^2)$ within a distortion level $D$, one needs $R(D)=\frac{1}{2}\log(\frac{\sigma^2}{D})$ bits. Suppose sources transmit their samples ignoring correlations. Since at one network use, the network can deliver $m$ linear combinations of source samples, $\frac{nN}{m}$ network uses are required. Moreover, since the noise variance is $\sigma^2$ but the allowed distortion level is $D$, roughly speaking, one needs at least $R(D)$ extra factor of transmissions. Overall,
\begin{equation}
\frac{nN}{m}\log(\frac{\sigma^2}{D})\nonumber
\end{equation}
 network uses are required. Note that, $k_1<<n$ and $k_2<<N$. Comparing this with the one of our proposed scheme (Theorem \ref{thm:main}) illustrates that, although our scheme is power efficient, it performs closely to an optimal scheme where only an informative subspace of source samples are transmitted.    
 
\item Note that, projection matrices $\bA_i$ should be known at receivers. Transmitting these matrices may cause a rate overhead in a practical sensing-communication scheme. In practical implementations, since this matrix is a random matrix, each source can use a pseudo random matrix generated (i.e., pseudo-random bit sequence (PRBS) generators) and transmit its starting seed, instead of transmitting the whole matrix. This would have  a negligible effect in the sensing-communication performance. Analyzing our proposed scheme with pseudo random matrices are beyond the scope of this paper.

\item Note that, the result of Theorem \ref{thm:main} is not deterministic and holds with high probability. There is always a small chance that, reconstructed signals have higher distortion than the allowed threshold.

\item As the last point, our proposed framework allows to have \textit{distributed sensing-communication}. That is, suppose there is one signal that receivers desire to reconstruct it. Each sensor has a low power budget to take all measurements and perform codings by itself. However, if several sensors (sources) are used, instead, each can take few measurements from the signal. By using our proposed scheme, receivers can reconstruct the original signal by using these measurements. In fact, our framework makes a virtual \textit{strong} sensor by using several \textit{weak} sensors.     
\end{itemize}

\section {Proofs}\label{sec:proofs}
In this section, we present proofs.

\subsection{Proof of Lemma \ref{lem:cascade}}

We first show that, $\bC_1\bG$ satisfies the RE condition with parameters $(\gamma\lambda_1^2,\alpha)$ where $\bG$ satisfies the RE condition with parameters $(\gamma,\alpha)$ and $\lambda_1$ is the minimum absolute eigenvalue of $\bC_1$. We have,

\begin{eqnarray}
\frac{1}{m}\|\bC_1\bG\bY\|_{l_2}^2 &=&\frac{1}{m} \nonumber
\bY^T\bG^T\bC_1^T\bC_1\bG\bY\\\nonumber
&=&\frac{1}{m}(\bG\bY)^T (\bC_1^T\bC_1)(\bG\bY)\\\nonumber
&\overset{\rm (1)}{\geq}& \frac{1}{m} \lambda_1^2 \|\bG\bY\|_{l_2}^2\\
&\overset{\rm (2)}{\geq}& \gamma\lambda_1^2 \|\bY\|_{l_2}^2\nonumber
\end{eqnarray}

Inequality (1) comes from a linear algebra lemma and inequality (2) comes from the definition of the RE condition for the matrix $\bG$.

In the following, we prove the second half of this Lemma. We assume that, for $\bY\in\cC(\cS;\alpha)$, $\bY\bC_2\in\cC(\cS;\alpha)$. Also, suppose $\lambda_2$ is the minimum absolute eigenvalue of the matrix $\bC_2$. Therefore, 

\begin{eqnarray}
\frac{1}{m}\|\bG\bC_2\bY\|_{l_2}^2 &=&\frac{1}{m} \nonumber
\bY^T\bC_2^T\bG^T\bG\bC_2\bY\\\nonumber
&=&\bY^T(\frac{1}{m}\bC_2^T\bG^T\bG\bC_2)\bY\\\nonumber
&=&(\bC_2\bY)^T(\frac{1}{m}\bG^T\bG)(\bC_2\bY)\\\nonumber
&\overset{\rm (3)}{\geq}& \gamma \|\bC_2\bY\|_{l_2}^2\\
&\overset{\rm (4)}{\geq}& \gamma\lambda_2^2 \|\bY\|_{l_2}^2\nonumber
\end{eqnarray}

Inequality (3) comes from the definition of the RE condition for the matrix $\bG$. Note that, in this part, we use the assumption that, for $\bY\in\cC(\cS;\alpha)$, $\bY\bC_2\in\cC(\cS;\alpha)$. Inequality (4) comes from a linear algebra lemma. This completes the proof.  

\subsection{Proof of Theorem \ref{thm:main}}

In this part, we prove Theorem \ref{thm:main}. 
Suppose $\bA_i\Phi$ satisfies the RE condition with parameters $(\gamma_1\lambda_1^2,\alpha)$ (i.e., $\bA_i$ satisfies the RE condition with parameters $(\gamma_1,\alpha)$ and $\lambda_1$ is the minimum absolute eigenvalue of the matrix $\Phi$. Then, use Lemma \ref{lem:cascade}). Also, say the matrix $\bG\bB\Psi$ satisfies the RE condition with parameters $(\gamma_2\lambda_2^2,\alpha)$ (i.e., in case 1, matrix $\bB\Psi_{\cS}$ satisfies the RE condition with parameter $(\gamma_2,\alpha)$ and $\lambda_2$ is the minimum absolute eigenvalue of $\bG^{\cS}$ where $\bG^{\cS}$ contains columns of $\bG$, corresponding to active sources. In case 2, $\bG$ satisfies the RE condition with parameters $(\gamma_2,\alpha)$ and $\lambda_1$ is the minimum absolute eigenvalue of the matrix $\Psi$). Hence, by using the LASSO decoder of equation (\ref{eq:lasso-decoder-1}) and the error bound of equation (\ref{eq:error-bound}), we have

\begin{equation}\label{eq:pf-part1}
\|\bmu-\tbmu\|_{l_2}^2\leq \frac{\delta_1}{\gamma_2^2\lambda_2^4}  \frac{k_2\log(N)}{m_2} \sigma^2.
\end{equation}

Hence,

\begin{eqnarray}\label{eq:pf-part2}
\|y_i-\tilde{y_i}\|_{l_2}^2 &\overset{\rm (1)}{\leq}& \|\bY-\btY\|_{l_2}^2\\
&\overset{\rm (2)}{=}& \|\Psi\bmu-\Psi\tbmu\|_{l_2}^2\nonumber\\
&\overset{\rm (3)}{\leq}& \lambda_3^2 \|\bmu-\tbmu\|_{l_2}^2\nonumber\\
&\overset{\rm (4)}{\leq}& \frac{\delta_1\lambda_3^2}{\gamma_2^2\lambda_2^4}  \frac{k_2\log(N)}{m_2} \sigma^2.\nonumber   
\end{eqnarray}

Inequality (1) is true for all vectors. Equality (2) uses the spatial correlation matrix property where $\bY=\Psi\bmu$. Inequality (3) uses a linear algebra lemma. Inequality (4) uses equation (\ref{eq:pf-part1}). 

Say $\sigma_u^2=\|y_i-\tilde{y_i}\|_{l_2}^2$. For the decoder of equation (\ref{eq:lasso-decoder-2}), we can write,

\begin{eqnarray}
\|x_i^t-\tilde{x_i^t}\|_{l_2}^2 &\overset{\rm (5)}{\leq}& \|\bX_i-\btX_i\|_{l_2}^2\nonumber\\
&\overset{\rm (6)}{=}& \|\Phi\bett-\Phi\tbett\|_{l_2}^2\nonumber\\
&\overset{\rm (7)}{\leq}& \lambda_4^2 \|\bett-\tbett\|_{l_2}^2\nonumber\\
&\overset{\rm (8)}{\leq}& \frac{\delta_2\lambda_4^2}{\gamma_1^2\lambda_1^4}  \frac{k_1\log(n)}{m_1} \sigma_{u}^2\nonumber\\
&\overset{\rm (9)}{\leq}& c \frac{k_1k_2\log(n)\log(N)}{m_1m_2} \sigma^2\nonumber  
\end{eqnarray}

where $c=\delta_1\delta_2\frac{(\lambda_3\lambda_4)^2}{(\gamma_1\gamma_2)^2(\lambda_1\lambda_2)^4}$.

Inequality (5) is true for all vectors. Equality (6) uses the temporal correlation matrix property where $\bX_i=\Phi\bett$. Inequality (7) uses a linear algebra lemma. Inequality (8) uses equation (\ref{eq:error-bound}). Inequality (9) uses equation (\ref{eq:pf-part2}). Having the distortion requirement $\frac{1}{n}\|\bX_i-\btX_i\|_{l_2}^2\leq D$ and $\cC_{\mbox{use}}=\frac{m_1m_2}{D}$ completes the proof of Theorem \ref{thm:main}.  

\section{Conclusions}\label{sec:conc}

In this paper, we proposed a joint source-channel-network coding scheme, based on compressive sensing principles, for wireless networks with AWGN channels (that may include multiple access and broadcast). A key idea of our proposed scheme is to reshape inherent redundancies among source samples (temporal or spatial) in joint source-channel-network coding to have a power/rate efficient sensing communication scheme. By characterizing the performance of the proposed scheme, we showed that, our proposed scheme
\begin{itemize}
\item (1) is power and rate efficient, that is, although it performs closely to optimal rates, it provides power efficiencies.
\item (2) has low decoding complexity where decoders are convex optimizations.
\item (3) is broadly structure independent, that is, sensing, coding and communication processes in sensors are application independent. 
\item (4) and has continuous rate-distortion performance, that is, if channel qualities change, the reconstruction quality will change in a continuous manner. 
\end{itemize}

In this framework, we perform joint source-channel coding at each source by randomly projecting source values to a lower dimensional space. We consider sources that spatially satisfy the restricted eigenvalue (RE) condition and may use sparse networks as well as more general sources which use the randomness of the network to allow mapping to lower dimensional spaces. The receiver uses compressive sensing decoders to reconstruct source signals.

%\begin{IEEEkeywords}
%Functional compression, Min-Cut theorem, graph Coloring, graph entropy.
%\end{IEEEkeywords}

\IEEEpeerreviewmaketitle

%\bibliographystyle{IEEEtran}
%\bibliography{IEEEabrv,refe}


\end{document}